# Study of the Use of Property Probes in an Educational Setting

Anton Risberg Alaküla[a], Niklas Fors[a], and Emma Söderberg[a]

a    Lund University, Sweden

**Abstract**

**Context**   Developing compilers and static analysis tools ("language tools") is a difficult and time-consuming task. We have previously presented *property probes*, a technique to help the language tool developer build understanding of their tool. A probe presents a live view into the internals of the compiler, enabling the developer to see all the intermediate steps of a compilation or analysis rather than just the final output. This technique has been realized in a tool called CodeProber.

**Inquiry**   CodeProber has been in active use in both research and education for over two years, but its practical use has not been well studied. CodeProber combines liveness, AST exploration and presenting program analysis results on top of source code. While there are other tools that specifically target language tool developers, we are not aware of any that has the same design as CodeProber, much less any such tool with an extensive user study. We therefore claim there is a lack of knowledge how property probes (and by extension CodeProber) are used in practice.

**Approach**   We present the results from a mixed-method study of use of CodeProber in an educational setting, with the goal to discover if and how property probes help, and how they compare to more traditional techniques such as test cases and print debugging. In the study, we analyzed data from 11 in-person interviews with students using CodeProber as part of a course on program analysis. We also analyzed CodeProber event logs from 24 students in the same course, and 51 anonymized survey responses across two courses where CodeProber was used.

**Knowledge**   Our findings show that the students find CodeProber to be useful, and they make continuous use of it during the course labs. We further find that the students in our study seem to partially or fully use CodeProber instead of other development tools and techniques, e.g. breakpoint/step-debugging, test cases and print debugging.

**Grounding**   Our claims are supported by three different data sources: 11 in-person interviews, log analysis from 24 students, and surveys with 51 responses.

**Importance**   We hope our findings inspire others to consider live exploration to help language tool developers build understanding of their tool.

## The Art, Science, and Engineering of Programming

**Study of the Use of Property Probes in an Educational Setting**

## 1 Introduction

Language tooling, like compilers and static analyzers, can easily become complex to develop. A professional compiler takes many person-years to develop and typically has to comply with complex semantic specifications. For example, the specification of Java version 8[1] is 788 pages long, and it contains many semantic rules that interact with each other. The language community has developed numerous tools over the years to assist with this complex activity. We have seen advances in areas such as language tool generation [15, 36] and full language workbenches [18, 37]. There has been a lot of progress in the development of language tool chains to enable faster development of language tools, but there are still more opportunities for improvements.

One activity in language tool development worthy of more attention is program comprehension, which underpins tool understanding, feature development, maintenance, and debugging. As we increase the level of abstraction and introduce more code generation into the workflow, the distance to the running code increases. Declarative specifications can be great as a way to separate the 'what' from the 'how', but when the specification is not doing what it should, it can be tricky to get insights into how to improve it. We believe that the nature of a declarative approach may introduce additional hidden dependencies (one of the cognitive dimensions [10]) that may decrease usability when things break down.

One approach to shed light on the "hidden" inner functionality of a language tool is to utilize so-called property probes [2]. Property probes, which have been realized in the tool CODEPROBER, provides a live, exploratory view into the functionality of a language tool. The aim with CODEPROBER is to assist during the development of language tools in a way that complements existing tools, with the goal to help explore and build understanding of language tooling. CODEPROBER is meant to be used both by students learning about building compilers and program analyzers, and by practitioners in industrial language tool development. CODEPROBER has been in active use in education and research for over two years now. However, we lack an understanding of how property probes are used in practice and what the users' experience of using them are.

In this paper, we present a mixed-method study on the use and user experience of CODEPROBER in an educational setting, focusing on students learning about compilers and program analyzers. We combine the results from 11 interviews, log analysis from 24 students, and survey results from 51 responses to find an answer to the following research questions:

- **RQ1** What is the user experience of using CODEPROBER in an educational setting?
- **RQ2** How is CODEPROBER used during the development of compilers and static analysis tools in an educational setting?
- **RQ3** How does the use of CODEPROBER compare to other tools used by students during the development process (debuggers, test cases, print-statements, AI, etc.)?

---
[1] https://docs.oracle.com/javase/specs/jls/se8/jls8.pdf. Accessed 2025-02-04.

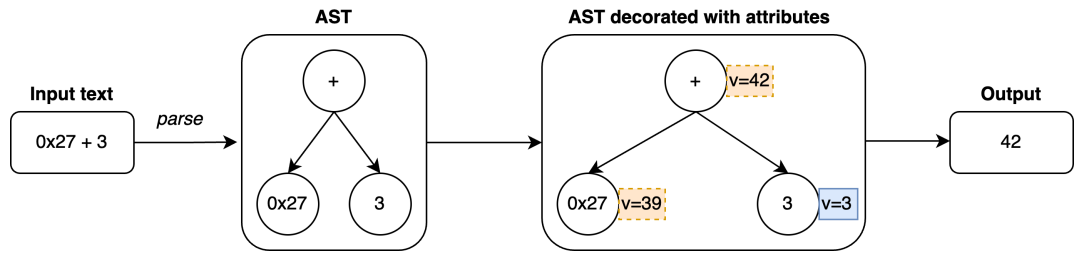

**Figure 1** Overview of how a RAG-based compiler works. The input text is parsed into an AST, and the AST is decorated with attributes. Attribute values are computed on-demand, meaning that nothing is computed until attributes are accessed. In the figure, the attribute v of the root node is accessed, which results in v and all its transitive dependencies being computed. Solid blue color indicates an intrinsic attribute, i.e., something known during parsing that requires no additional computation. Dashed orange color indicates something that is computed.

We find that the students find CodeProber to be useful, and they make continuous use of it during the course labs. We further find that the students in our study seem to partially or fully use CodeProber instead of other development tools and techniques, e.g., breakpoint/step-debugging, test cases, and print debugging.

The rest of the paper is structured as follows. We start by giving some background into how language tooling is built, specifically focusing on Reference Attribute Grammars [14] which CodeProber works with (Section 2), before we give an introduction into CodeProber's features (Section 3). We then introduce the overall design of the study (Section 4), followed by the method and results for the interview part (Section 5), log file analysis part (Section 6), and survey part (Section 7). Finally, we discuss the results in light of our research questions (Section 8), before we discuss related work (Section 9) and conclude (Section 10).

## 2 Language Tools with On-Demand Evaluation

In this paper, we refer to compilers and static analysis tools as "language tools". They have similar overall goals: transform source code into some desired output. The output can be machine code, a list of diagnostic messages, the result of running test cases, etc. The implementation of a language tool can be pass-based, where values needed internally in the tool are computed in separate passes, e.g., one pass for name analysis and another pass for type analysis. Alternatively, the computation can be on-demand, e.g., some computed value is asked for, and then only that value is computed, including all the values it transitively depends on.

In this section, we provide a high-level overview of Reference Attribute Grammars (RAGs) [14], an approach for building language tools that compute values on-demand. RAGs are an extension of Attribute Grammars [19]. RAG-based tools start by parsing source code into an abstract syntax tree (AST). They then associate functionality called *attributes* with AST nodes. Attributes may depend on other attributes, and can compute values on-demand. This lets the developer specify a full language tool as a set of smaller attributes that depend on each other. For example, the entry point of the compiler may access an attribute errors representing compile-time errors. When errors is accessed, errors and all attributes needed to compute errors, such as type attributes, are computed automatically. The computed attributes are also memoized for subsequent accesses.

Figure 1 provides an illustration of a RAG-based language tool in the form of simple calculator language. The language supports additions and two kinds of integers: base-10 and base-16. The desired output for this language is the base-10 value of the program. For example, for the input 0x27 + 3, the output should be 42. The computation is defined as an attribute v that has different definitions depending on the node type. For example, the definition for addition nodes accesses v on its children and adds them together. The program output is then computed by accessing the attribute v on the root node (which might trigger more attributes to be computed).

The on-demand aspect of RAGs enables a small subset of attribute values to be accessed and usually computed very quickly. For example, the performance evaluation by Alaküla et al. [2] shows that some selected attributes finish evaluating in $1-30$ milliseconds. The exact time it takes depends on the chosen attribute's complexity, but it often finishes in less than 100 milliseconds. In general, a response time within 100 milliseconds in an interactive system appears instant to the user [27]. The low response times of RAGs enable quick, interactive exploration of the inner workings of the language tool, which can be of great help during the development process.

**Debugging RAGs**  In order to debug and build understanding of a RAG-based tool, it is beneficial to be able to explore each attribute individually. Accessing and evaluating these internal properties is possible to do with traditional debugging tools and techniques, but we believe it can be inconvenient to do so. In this section, we give examples of the problems that may arise, especially when working with language tools for non-trivial languages.

Assume a language tool developer wants to inspect a specific attribute like v in Figure 1 using print debugging. They would have to: 1) Add a print statement to the language tool specification, 2) Rebuild the language tool, 3) Run the tool with an example input file, and 4) Filter the output to find the line corresponding to the node of interest. Steps 1 and 2 need to be done for every new information the developer wants to extract, and build times can be long for larger language tools. Additionally, step 4 can be a significant hurdle when exploring larger input files, as the same attribute may be invoked many times for different AST nodes. Step 4 can be mitigated by making the print statement conditional. However, this might require some non-trivial conditional expression. Similar issues arise with traditional breakpoint/step-debuggers, where defining a conditional breakpoint for a particular AST node might be non-trivial when there are many nodes of the same type.

Print debugging and breakpoint/step-debugging are viable options for exploring attributes on an AST. However, we believe it is more convenient to freely explore attributes without rebuilding the language tool and to use the input text to find nodes of interest. This has been realized in the tool CODEPROBER [2] that supports property probes, which we will describe in the next section.

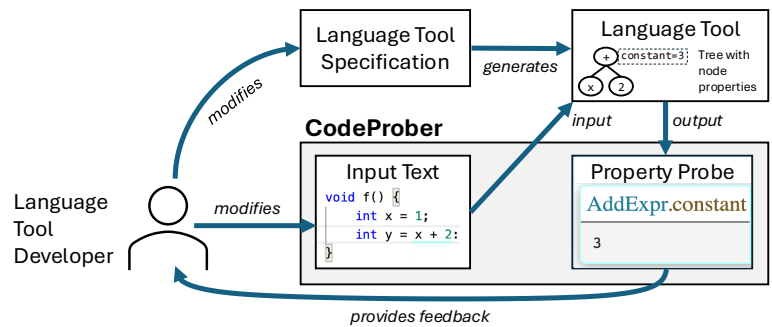

**Figure 2** Overview of CODEPROBER. The language tool developer has created a property probe of the expression x + 2 in the input text. The property constant computes the compile-time constant of that expression. The developer can either change the input text or the language tool, and the probed result is automatically updated.

## 3 CODEPROBER

This section gives a brief introduction to CODEPROBER [2] and its features.[2] CODE­PROBER allows for an interactive exploration of intermediate results in language tools. The results are live and automatically updated when the input text (typically code) or the language tool changes. An overview of CODEPROBER is shown in Figure 2.

CODEPROBER simplifies accessing and exploring values that a language tool computes. CODEPROBER requires that the language tool parses input text into a tree with node properties that can be accessed. The language tool developer can then create *probes* via the input text in CODEPROBER to access these properties. The requirement of node properties maps very well to attributes described in the previous section. However, property probes are a general technique that can be applied to other kinds of language tools as long as they fulfill the requirements, i.e., they associate properties with nodes in a tree.

**Basic Usage** Basic usage of CODEPROBER starts with the language developer writing the input text they wish to explore into the CODEPROBER editor, and then they right click to create a probe. Figure 3 shows the process of creating a probe in the Java compiler ExtendJ [6]. In it, the developer started by writing a small Java program. Then, they create a probe for the compile-time constant value of an addition expression. From this point, if the developer makes any change to the Java program, or if they update the underlying Java compiler, then the probe will display updated values. The developer can then continue to create probes for other properties, for example, byte code generation, without requiring recompilation. This is different from e.g. print debugging, where the developer has to modify source code, compile and run for each new piece of information they wish to extract.

This ability to explore properties, together with the liveness features, is central to the experience of using CODEPROBER. Normal probes, as shown in Figure 3, is a core way of interacting in CODEPROBER. However, several more ways of interaction

---

[2] A video demonstration of CODEPROBER: https://youtube.com/watch?v=lkTJ4VLOxtY.

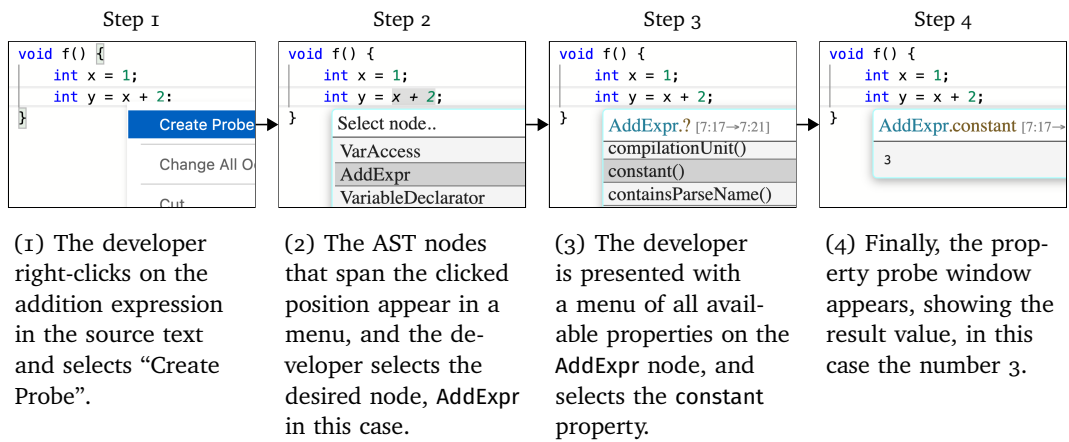

(1) The developer right-clicks on the addition expression in the source text and selects "Create Probe".

(2) The AST nodes that span the clicked position appear in a menu, and the developer selects the desired node, AddExpr in this case.

(3) The developer is presented with a menu of all available properties on the AddExpr node, and selects the constant property.

(4) Finally, the property probe window appears, showing the result value, in this case the number 3.

**Figure 3** Steps to create a probe for the constant property of the addition expression x + 2.

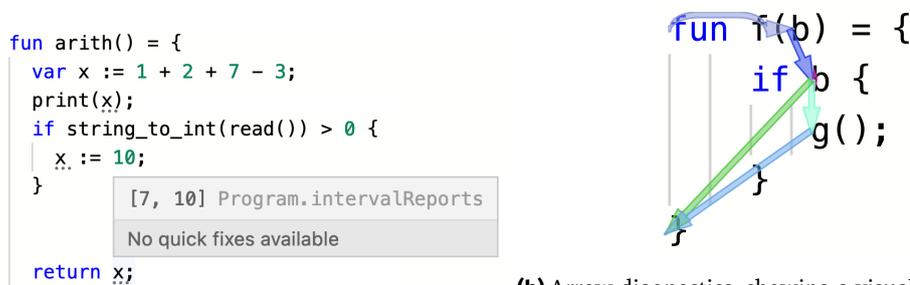

**(a)** Squiggly line diagnostics

**(b)** Arrow diagnostics, showing a visualization of the control-flow graph of a function.

**Figure 4** Screenshots of diagnostic contributions in CODEPROBER. The most common kind of diagnostic is "squiggly lines", and they are shown in three locations in Figure 4a. The user is hovering the last location (x variable on the last line), and the popup shows the variable's interval value at that point in the program. In this case, the value of x on the last line must be within the interval of [7, 10].

are supported. Some of these are illustrated in the following list using a teaching language called TEAL. See also Section 4.2.3 for more information about TEAL.

- *Diagnostic contributions* allow probes to contribute diagnostics to the text editor, in the form of "squiggly lines" (Figure 4a) and arrows (Figure 4b). This can be used for example to present semantic issues or render a control-flow graph.
- *AST probes* display part of the AST in graphical form. This can be seen in Figure 5.
- *Search probes* support finding a set of nodes that pass some predicate, and evaluate an attribute for all of them simultaneously. A search probe can be seen in Figure 6.

Currently, we have implemented support for language tools specified with the meta-compilation system JastAdd [7]. JastAdd combines reference attribute grammars with object- and aspect-orientation. For an introduction to JASTADD, see [13]. However, there is an interface that other kinds of language tools can implement to integrate with CODEPROBER.

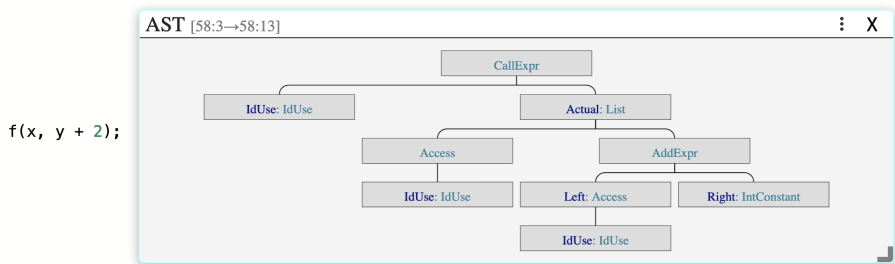

**Figure 5** AST probe showing the AST of a function call. More probes can be created by clicking on individual nodes in the AST.

```
 3
 4
 5
 6
 7
 8  fun arith() = {
 9    var x := 1 + 2 + 7 - 3;
10    print(x);
11    if string_to_int(read()) > 0 {
12      x := 10;
13    }
14    return x;
15  }
16
17
18
19
20
21
```

FunDecl.*.interval?isAccess [8:1→15:1]
Found 3 nodes
Access [10:9→10:9]
.interval [10]
[7, 7]

Access [12:5→12:5]
.interval [12]
[7, 7]

Access [14:10→14:10]
.interval [14]
[7, 10]

**Figure 6** Search probe that finds all nodes in the function (FunDecl) where property isAccess is true, and opens a nested probe for property interval on them. The user is hovering the middle Access node, which causes the corresponding span in the text editor to be highlighted. (Note that the middle result of [7, 7] is correct, it displays the interval value prior to the assignment).

## 4 Study Overview

In this section, we provide an overview of our study along with a description of the context in which the data for the study was collected. We designed a mixed-method study to gain a deeper understanding of how property probes are used in practice via the CodeProber tool, and the user experience of using CodeProber in an educational setting with students learning about compilers and program analysis. The objective of the study is to address the research questions defined in Section 1.

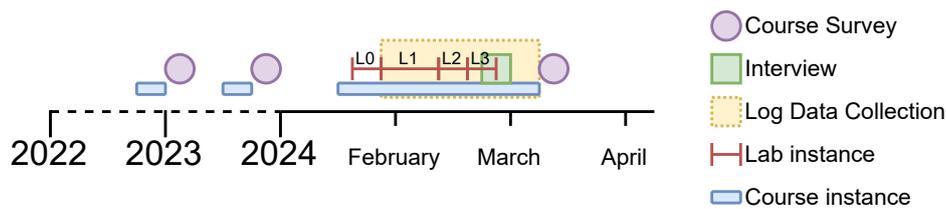

■ **Figure 7** Overview of when the different parts of the study took place. The four red lines labeled "L0" to "L3" represent the planned durations of Lab 0 to Lab 3 in the program analysis course. Late lab submissions were allowed, which is why the log data collection proceeded after the end of Lab 3.

## 4.1 Study Objective and Design

The study is composed of three parts; interviews with students, analysis of logs from students using CodeProber, and a survey sent to students after they used Code-Prober as part of course labs. An overview of when the different parts took place is seen in Figure 7.

Each part helps address one or more of the research questions. The interview questions are designed to help answer each research question (**RQ1**, **RQ2**, and **RQ3**). The log analysis tells us how CodeProber is used, which helps answer **RQ2**. The course survey contains a question about the "effectiveness" of CodeProber, which helps in answering **RQ1**. Triangulation, in the context of user studies, refers to investigating a phenomenon from at least two different perspectives [34]. Perspectives can mean making observations at different points in time, using different techniques, observing different groups of people, etc. The main idea is that by observing the same phenomenon from multiple perspectives, it is possible to draw conclusions with more confidence. We hope that our study design, combining three parts with different methods, will help to answer our **RQs** with better confidence.

The details of each of the study parts, along with their results, are presented in Section 5, 6, and 7, respectively. For the rest of this section, we will focus on describing the context of the study.

## 4.2 Educational Setting

The data collection for this study is carried out in connection to two university courses taught to engineering students typically specializing in computer science; one course on program analysis and another course on compilers. Both of these courses have integrated the use of CodeProber in the practical work in the course. Our data collection is focused on the Spring 2024 instance of the program analysis course, but we also include survey responses gathered for the Fall 2023 instance of the compilers course and also the earlier Fall 2022 instance of the program analysis course.

It should be noted that students who take the program analysis course have often taken the compilers course prior to it (or another compilers course) and at least two other programming courses. Of the 31 students that finished the 2024 instance of the program analysis course, only 2 had not taken any prior compiler course. 2 had taken a compiler course at a different university, and 3 had taken the compiler course at our university before CodeProber was introduced. 24 of the 31 students had therefore seen CodeProber before, but we do not have any concrete data on whether they made use of it, as usage was optional.

In the following subsections we describe the student population of the program analysis, as well as the two courses in more detail.

### 4.2.1 Student Population

The average student in the program analysis course is in their 4th year of a 5-year computer science and engineering program[3] at Lund University. In the 2024 instance of the course, there were 31 students that finished the course, and 27 of these were from the computer science and engineering program. The remaining 4 students consisted of 2 PhD students and 2 students from a different engineering program.

The first three years at the computer science and engineering program contains 7 mandatory programming-related courses, totaling 44.5 ECTS credits. These courses cover topics like object-oriented programming (Scala and Java), concurrency, agile software development (including test-driven development) and functional programming. There are also several labs where traditional debuggers are used.

The students in the program analysis course are ideal candidates for our user study since they develop language tools in an educational setting, and since CodeProber has been incorporated in the course. They further have experience from debugging in previous courses, which gives them a good foundation for reasoning about CodeProber and comparing it with alternatives.

As can be seen in Figure 7, we interview the students at the end of the last lab in the course. This is so that they have time to build as much experience with CodeProber as possible, and they still have that experience fresh in their heads when we interview them.

### 4.2.2 The Compilers Course

In the compilers course (7.5 ECTS credits), students learn how to create a compiler from scratch. Over a series of 6 labs they create a basic C-like compiler. They do this incrementally, starting with scanning and parsing, and eventually adding code generation.

CodeProber is introduced to the students at the end of lab 3, at the same time as they start writing their own RAG attributes. Then for the remainder of the labs, they are told that they should implement test cases for everything they do, but they are also welcome to use CodeProber if they want.

At the end of the lab sessions we usually ask the students how they approached solving the lab. We estimate that roughly two thirds of them mention using CodeProber in the 2023 instance of the course. This estimate, while not based on hard data, is also reflected in the course evaluation for the compilers course (Figure 9c), where just over two thirds of the students (15 of 22) "agree" or "fully agree" that CodeProber is "effective".

---

[3] Full name: "Master of Science in Engineering, Computer Science and Engineering".

Study of the Use of Property Probes in an Educational Setting

### 4.2.3 The Program Analysis Course

In the program analysis course (7.5 ECTS credits), the students implement several different kinds of program analyses on top of an existing compiler in a language called TEAL ("Typed Easily Analysable Language"). TEAL is a gradually typed imperative language. An example of TEAL code from the labs is included in Figure 4a. At the start of each lab, the students are provided with a working compiler and instructions for what to add. The provided compilers are all implemented using JASTADD. In the 2024 course instance, the following labs were included:

**Lab 0: Introduction to JASTADD and CODEPROBER** A relatively short lab, specifically designed to assist students with no prior experience with JASTADD. In this lab, they implement simple type checking. No log data was collected from this lab.

**Lab 1: Type Inference** In this lab, the students implement monomorphic type inference, based on collecting and solving constraints.

**Lab 2: Dead-Assignment Analysis** In this lab, the students mainly implement dead-assignment analysis, i.e., given a declaration or assignment, compute whether the assignment is unnecessary.

**Lab 3: Interval Analysis** In the final lab, the students implement interval analysis, i.e., for all integer variables, compute its possible range of values. An example from this lab is included in Figure 4a.

The size of the labs vary significantly, both in terms of how much code is handed out and how much code is required to solve them. The handout code varies from 3000 to 4000 lines of code (not counting tests). The solutions to the labs vary from 100 to 600 lines of code. A snapshot of the handout code for all labs is available on Zenodo [31].

Each lab contains a small set of example files that can be opened in CODEPROBER. The version of CODEPROBER used in the labs is preconfigured to extract some diagnostic information from the students tools. This diagnostic information is usually presented as hoverable squiggly lines or dots, such as the dots seen in Figure 4a.

The students are encouraged to write tests, and use of CODEPROBER is in theory optional. However, examples are often given in terms of CODEPROBER, and the TAs often ask to see functionality via CODEPROBER, so in practice all students actively use it to some degree.

## 5 Interviews

This section presents the method and the results of the semi-structured interviews carried out with students taking the Spring 2024 instance of the program analysis course described in Section 4.

### 5.1 Method

Below, we describe how we designed and executed the interview study, i.e., how we collected data in the study and how we analyzed that data.

### 5.1.1 Data Collection

As a first step of the data collection for this interview study, we designed the interview protocol in connection to the research objective. After testing the interview protocol with a pilot interview and refining the protocol, we continued to recruit participants and to carry out the interviews. See below for further details.

**Designing the Interview Protocol**   The interview protocol includes the following parts:

1. **Warmup** We ask background questions about the participants experience and skill, and basic questions about CodeProber. Some of these questions do not directly relate to a research question, but help us get to know the participant, and helps the participant to start thinking more about CodeProber.
   For the questions about skill, Peitek et al. [29] found that one of the best indicators of "programming efficacy" is to ask the programmer to rate themselves compared with their peers. Therefore, in the background form we asked the participants to rate their programming skills in comparison to their classmates on a scale of $[-2, 2]$, where $-2=$"Much Worse", $-1=$"Worse", $0=$"Identical", $1=$"Better", $2=$"Much Better".

2. **Workflow** We ask a set of questions relating to how the students approach the labs. What editor do they use, do they write test cases, when is CodeProber used, etc. This section relates to **RQ2** and **RQ3**.

3. **Scenarios and Tools** We ask how much the student uses different development tools during different scenarios of working in a code base. This discussion is driven by them filling in a table of how much they use each tool when developing language tools (mostly in the program analysis course). One axis of the table contains development tools, and the other axis contains development scenarios. The included tools are "Print debugging", "Breakpoint/step debugging", "Test cases", CodeProber and "AI" (Copilot/ChatGPT/etc.). The scenarios are "Developing a new feature", "Developing understanding of a codebase" and "Fixing a bug". There is some overlap between the scenarios, but they roughly map to the middle, beginning and end of the lab, respectively. Finally, we ask them how much they *like* using each development tooling. This section relates to **RQ1** and **RQ3**.

4. **Likes and Dislikes** Finally, we ask the participants what they like most and least about CodeProber. We also ask if there are any features they would like to add. This ends up being somewhat a repetition of what was said in the previous two sections, but it gives the interviewee a chance to highlight which things matter most. This section relates to **RQ1** and **RQ2**.

After designing the main interview protocol we performed a pilot study. Afterwards a few minor things were adjusted. The end result is an in-person interview that takes about 50 minutes to perform. The interview is a mix of open questions and some forms for the interviewee to fill in. The full interview design is available in Appendix A, B, C and D. It is also available on Zenodo [1], together with results from the log file analysis.

**Study of the Use of Property Probes in an Educational Setting**

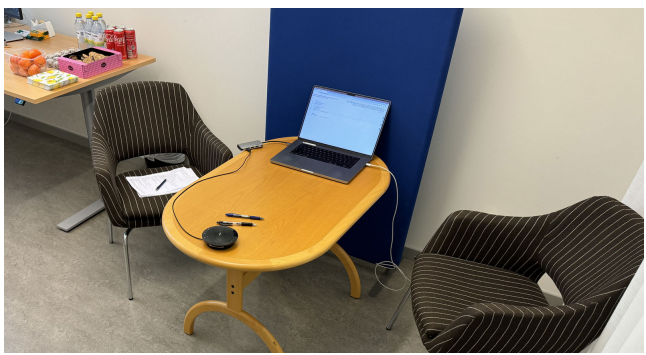

**Figure 8** Room used during the interviews.

**Recruiting Participants**   We attended one of the course lectures and presented our intention to perform this study. An outline of the interview was shown, as well as their rights and expected reward (a small take-home gift). We also stressed that the interview would not measure or benchmark the students in any way. We hoped this would make students more eager to apply, and it seemed to have worked as 9 of them applied (out of 31). In addition, we interviewed 2 teaching assistants that either are or were involved in the course, but are not part of the team working on CodeProber. The pilot study was carried out with one of the TAs. In total, we interviewed 11 people.

**Execution of the Interviews**   The first and second author of this paper met each interview subject in turn. We sat around a table that had a laptop running CodeProber, to be used as demo/reference if necessary. One led the interview and the other took notes. A picture of the room used during all interviews is visible in Figure 8. The interviews were recorded after informed consent from participants. The laptop had active screen- and voice-recording throughout the interview. In addition, a phone was used to record voice for redundancy.

#### 5.1.2 Data Analysis

The interviews provided two kinds of data: the forms and audio recordings. The form data was relatively small, and was manually entered into a spreadsheet. The audio required more processing. First, we transcribed all recorded interviews using OpenAI Whisper [30]. Then, we did a manual inspection over all the transcripts and fixed any errors from the model. This resulted in 1770 lines of text being added and 2434 lines being removed. The corrected transcriptions are in total 11607 lines of text. Finally, we performed coding.

**Coding the Transcripts**   A coding scheme was developed to help with extracting information from the interview transcripts. The goal of this coding scheme was to extract common themes, with regard to the use and experience of using CodeProber and other development tools, as formulated in the research questions. The first and second author of this paper independently extracted a coding scheme from the same transcript. Then they met, discussed the result and merged the two coding schemes into one. This process repeated three additional times, until the coding scheme was relatively stable. Then, the scheme was applied to all transcripts. The final coding was reviewed by the third author. The final coding scheme can be found in an artifact on Zenodo [1].

### 5.2 Results

During the interviews, our participants filled in three forms focused on their experience and skill, feature usage, and development techniques and their use in different scenarios. We will present the results from these three forms before we move on to present the themes constructed from the analysis of the remaining interview data. The data from the forms is split into two groups: *students* and *TAs*. The student group has 9 participants, and TAs has 2. This split is done since the TAs background and relation to the course differs significantly from the average student.

#### 5.2.1 Participants Experience and Skill

The median student we interviewed had 6 years programming experience and was in their 4th year of university studies. All but one of the students had previously taken the compiler course at our university. One interviewee was a PhD student, and the rest were students in the computer science and engineering program (see Section 4.2.1).

For the self-assessed skill, we got responses of 0 and 1, with the average response being 0.45. With little variation in the result for the skill self-assessment, we decided to not split results based on skill.

#### 5.2.2 CodeProber Feature Usage

The feature usage form contains a list of features inside CodeProber. The interviewees were asked to fill in how much they use each feature, on a scale from "Never" to "Very Often". They could also answer with "x" if they did not know said feature existed. The results are presented in Table 1.

The table shows a few clear winners in terms of features. Creating probes, looking at their outputs, and performing live updates are all very common. These features also happen to be the ones that are described in detail in the lab instructions. Some other features are less used or less well known, such as search probes and tracing (showing what attributes, with their intermediate values, an attribute depends on).

#### 5.2.3 Techniques and Scenarios

The technique and scenario form consists of two parts. First, some questions of how much different development techniques are used for different scenarios. The context here is developing language tools, and most students (N=5) answer it solely based on the experiences in the courses. Second, the interviewees are asked to rate how much they *like* using each technique. The results are shown in Tables 2 and 3.

#### 5.2.4 Interview Themes

Here we present the main themes constructed from analyzing the interview transcripts. In cases where individual quotes are used, the participants name is presented as P*X*, where *X* is an integer in the range of [0, 10]. P0 and P10 are teaching assistants, the

**Study of the Use of Property Probes in an Educational Setting**

■ **Table 1** Average usages of CODEPROBER features. Feature names are shortened to fit the page, see Appendix C for full names. Usage rate is reported on a scale of $[0, 5]$, where 0=Never, 1=Very Rarely, 2=Rarely, 3=Sometimes, 4=Often, 5=Very Often. In case the interviewee had not heard about the feature, they could also answer "x", which is shown as a separate column.

| Feature | Students Use | Students #x | TAs Use | TAs #x |
|---|---|---|---|---|
| Creating probes from text (e.g Figure 3) | 4.67 | 0 | 4.50 | 0 |
| Creating probes from node references | 3.88 | 1 | 5.00 | 1 |
| Inspecting probe outputs | 4.56 | 0 | 4.50 | 0 |
| Hovering AST node references | 4.00 | 0 | 2.00 | 0 |
| Inspecting the AST (e.g. Figure 5) | 2.39 | 0 | 1.50 | 0 |
| Creating nested probes ('▼') | 3.75 | 1 | 4.00 | 0 |
| Using Minimized probes | 1.17 | 3 | 2.00 | 1 |
| Using arrows, showing e.g. the control-flow graph | 2.61 | 0 | 2.00 | 0 |
| Looking at/conveying information in squiggly lines | 4.44 | 0 | 3.50 | 0 |
| Liveness from text updates | 3.78 | 0 | 4.00 | 0 |
| Liveness from rebuilding the compiler | 4.56 | 0 | 4.50 | 0 |
| Using the "Stop" button for long-running probes | 2.00 | 2 | 1.00 | 1 |
| Search probes (e.g. Figure 6) | 2.00 | 8 | 1.00 | 1 |
| Tracing | 1.20 | 4 | 1.50 | 0 |

■ **Table 2** Average usage of tools for different development scenarios. Scale is $[0, 5]$, where 0=Never, 1=Very Rarely, 2=Rarely, 3=Sometimes, 4=Often, 5=Very Often. Some names have been shortened to fit the table, full table is available in Appendix D. Note that these numbers are in the context of developing language tooling; they do not apply for software development in general.

| Tool | New feature Students | New feature TAs | Understanding code Students | Understanding code TAs | Bugfixing Students | Bugfixing TAs |
|---|---|---|---|---|---|---|
| Print debugging | 1.44 | 3.50 | 1.11 | 0.5 | 2.33 | 3.50 |
| Breakpoint/step | 0.33 | 1.50 | 0.11 | 3.5 | 1.22 | 3.00 |
| Test cases | 2.67 | 4.50 | 1.33 | 0.5 | 2.50 | 4.00 |
| CODEPROBER | 4.44 | 2.50 | 4.44 | 0.0 | 5.00 | 3.25 |
| AI | 0.78 | 1.75 | 0.22 | 0.0 | 0.44 | 0.50 |

■ **Table 3** Average responses to how much the interviewee likes using each development technique. Scale is $[1, 5]$ where 1=Strongly dislike, 2=Dislike, 3=Neutral, 4=Like and 5=Strongly like. Tool names have been shortened to fit the table, full table is available in Appendix D.

| Tool | Students | TAs |
|---|---|---|
| Print debugging | 3.67 | 2.00 |
| Breakpoint/step | 3.89 | 3.75 |
| Test cases | 3.11 | 3.50 |
| CODEPROBER | 4.78 | 4.25 |
| AI | 2.33 | 2.50 |

rest are students in the program analysis course. Any quotes that were originally in Swedish have been translated to English.

We believe the interviews achieved a degree of data saturation [11]. The later interviews mostly repeated themes that earlier interviews had brought up, which makes us believe that our sample size is good enough to make some meaningful observations. We use the syntax "(N=NUM)" below to indicate how many (NUM) of the 11 participants mentioned a given theme.

**Theme: Liveness** All (N=11) participants mentioned that the *liveness* is a positive aspect of CodeProber. Liveness comes in two forms; changing the input text inside CodeProber, and updating the language tool being explored. Some interviewees (N=5) mentioned that they rely more on the second kind of liveness, because they know of a specific example that produces incorrect behavior, so there is no need to keep changing the text. They instead keep working on their compiler until the probes display the expected output. This way of working has some similarities to test-driven development. Some related quotes:

> P3: *"Because it's often that you are trying to debug something. You have a piece of code that should produce an error, but there is no error. So you change your own code and recompile and see, is there an error? No, not now either."*
> P4: *"Often you have an example you create. And then you use hot reloading to see, to make it work."*
> P5: *"[talking about changing the text] ...not that often actually, it feels like I often have a specific example I want to look at. Rebuilding the compiler and seeing the probes update live, yes quite often."*
> P6: *"Most of the time I write my example and then run CodeProber"*

Still, everybody relies on the first kind of liveness as well, which can be seen in Table 1.

**Theme: Exploration** All (N=11) participants mentioned some form of exploration when talking about CodeProber. Liveness, mentioned earlier, is one form of exploration, in that it enables the developer to explore which combinations of input leads to what output. There is also exploration in terms of listing which AST nodes exist, which properties are available to use, and how the different properties link the AST nodes together. Some related quotes:

> P2: *"many times we have searched how to jump through the AST"*
> P5: *"CodeProber is nice because you like do not have to set up anything, you just click around."*
> P7: *"[when asked about what they like most about CodeProber] it is how you can step through if you want to find.. like, for example that you can explore a bit. [..] If you did not use CodeProber it would be very difficult to find what methods to use."*
> P8: *"It [CodeProber] is very good in how you can visualize things. And easily understand, if you have a node here, then I can see all functions that can be used"*

Study of the Use of Property Probes in an Educational Setting

**Theme: Freezes and Crashes**   By far the most common negative feedback relating to CodeProber are about freezes and crashes, with all but one mentioning having issues (N=10). Most of the mentioned issues relate to a specific lab in the course. In this lab, the students implement interval analysis, i.e. they should find out what interval a variable's value may have a given point in the program. To handle loops, their analyses must run iteratively until the analysis values converge. To make sure that the analysis always finishes in a reasonable time they must additionally implement *widening*, i.e. overapproximate the interval values after a certain number of iterations.

During development most students had some bugs in their implementation, such as not implementing widening correctly, and this could cause their analysis to get stuck in an infinite loop. In some scenarios, this was not presented clearly in the CodeProber UI, causing some (N=4) to state that they can't always trust the values in the CodeProber UI, because they might be stuck. CodeProber tries to recover from this stuck state, but did so incorrectly, which meant that even when CodeProber became responsive again, it could be that the values shown in the UI aren't accurate. Some related quotes:

> P4: *"it's maybe your compiler that is wrong, but sometimes it's that CodeProber get stuck and you have to restart it."*
> P5: *"I have no idea why it crashed, because later on it didn't crash at all"*
> P8: *"I don't know if the error is because there is an error in my code, or that something froze"*

**Theme: Reduced Use of Other Tools**   While CodeProber isn't a direct replacement for any other tool, it does seem to lead to reduced use of the other tools. A majority of the participants (N=9) mention that CodeProber has partially or fully replaced testing. Some related quotes:

> P2: *"sometimes I am content with seeing that it works as I want in CodeProber."*
> P4: *"if I have CodeProber then I write fewer tests because in a way I have verified it by hand"*
> P7: *"when you are finished with something and are going to write a lot of test cases when you already know that it works, thats annoying"*

This reduction in testing is not positive, as CodeProber does not help prevent regressions. All but one of the participants (N=10) mention that one of the positive aspects of test cases is that it helps prevent regressions. A majority (N=6) of interviewees also mention that they dislike the process of writing test cases. We believe it is important that the students learn to work with tests more, but also understand some of their rationale for not writing tests here. It is "just" code for a lab after all, and perhaps they would be more open to writing tests for a long-term project.

A majority of participants (6), mentioned using print debugging more outside the course. This is in part due to the nature of the lab assignments. In the labs, the student code could run iteratively until values converged. If print statements were added there, there could be thousands of log entries, reducing the efficiency of printing.

Similarly, some (N=4) mentioned using debuggers more outside the course. This is in part because in the course the students work with JastAdd, and debugging support is not very strong for it. JastAdd generates Java code, and this can be stepped through with any standard Java debugger. However, the developer would have to step through some internal evaluation code that JastAdd generates, which negatively impacts the experience. CodeProber seems to partially substitute both printing and stepping for the students. Some related quotes:

> P3: *"CodeProber has kind of replaced print debugging in these courses. Because you can, without writing a lot yourself, just open and look at it. ['it' being an attribute value that would otherwise be printed]"*
> P4: *"Breakpoint debugging is easier if you are in a project that doesn't have a lot of JastAdd caching code and such, but is more straight forward. There I use it more. And I use print-f debugging more then as well."*
> P8: *"CodeProber [..] I compare it a lot to a breakpoint/step-debugger. Just because it makes it a lot easier to understand."*

**Theme: Reasons for Using AST View**   Some participants (N=4) mentioned making use of the AST view (shown in Figure 5). The reasons for doing this include trying to understand the structure of the AST, seeing which children belong to which node, and more. Some related quotes:

> P1: *"I needed to find what one of the children nodes was called."*
> P2: *"It feels like I did it a lot in the beginning, but maybe less and less.. [..] when you have an understanding of how it is built, then you do not need to look at it."*
> P5: *"AST View, it is mostly when I've forgotten which attributes exists and I don't feel like looking at the generated Java code, or the pre-written source code. [..] You can see them [the AST nodes] and it is easy to see the methods"*

Even among the participants using the AST view, usage was quite low. In Table 1, the average usage of the AST view is 2.39 ($\simeq$"Rarely"). Most (3 of the 4) participants that mentioned using the AST view also mentioned using alternative methods. For example, to understand the structure of the AST, it can be convenient to look at the definition of the AST structure (abstract grammar) instead. Only one person said that they use the AST view "Often", and this is the only participant that did not previously take the compilers course.

We initially added the AST view in response to students in the compilers course requesting it. They requested it after having seen a similar feature in another tool called DrAST [24], and mentioned that the graphical view helps to build understanding of the structure of the AST. However, once that understanding is achieved, then the AST view might not be as helpful anymore. This matches what some participants said. A possible takeaway from this is that the usefulness of an AST view is inversely proportional to the users experience level. Therefore, when building a tool meant to be used by students, or a tool to help introduce people to a larger codebase, consider adding an AST view. When building a tool for more experienced users, then it might not be as important to add.

**Study of the Use of Property Probes in an Educational Setting**

**Theme: Continuous use of CodeProber**   When asked about their workflow throughout the labs, a majority of participants (N=8) talked about an iterative process where they switch between writing code and checking the result inside CodeProber. Some (N=4) also said their usage or CodeProber goes up or down depending on where in the labs they are. Some related quotes:

> P4: *"[..] when you get into the middle [of the lab] the usage goes down, and then towards the end and beginning you use it [CodeProber] a lot."*
>
> P7: *"more in the end [of the lab] you do smallfixes and such, then you go into Code-Prober to see if things actually changed or not. So more coding in the beginnning, and then more CodeProber in the end."*
>
> P8: *"I would say that in the middle [of the labs] you spend more time looking at code. Because I have [..] identified the parts I need to work on in CodeProber. [..] And towards the end you go back [to CodeProber] to verify if that works."*
>
> P9: *"Towards the end it is a lot more CodeProber. [..] It is about trying to find why things go wrong. I go into the code, make a small change, recompile and check inside CodeProber again."*

We asked the participants for a rough estimate of how much of the total lab time they spend inside CodeProber. Both the average and median response of the 7 answers was 30 % inside CodeProber. The lowest individual answer was 5 % from one of the TAs. They said that the reason for this is that they started using JastAdd and building compilers before CodeProber existed, so they have gotten used to working with test cases and print debugging instead.

## 6  Log File Analysis

In this section, we present the method and the results from the log file analysis part of the study.

### 6.1  Method

For the 2024 instance of the program analysis course, we modified CodeProber to log user interactions into a file on the users machine. Here, we describe the design of these log files, how we collected the data, and how the data was analyzed.

#### 6.1.1  Data Collection

The log files generated by CodeProber contain lists of JSON objects on the following form:

```
{"s":SESSION, "t":TIME, "d":{ "t":TYPE } }
```

SESSION is an ID that is randomly generated every time CodeProber is started. TIME is a timestamp. TYPE is the type of event. Depending on the type of event, the object containing TYPE can contain more fields.

■ **Table 4** The number of collected events per lab, and the number of repositories those events were collected from. Lab 1 was performed in groups of two, so "12" represents 24 people. Labs 2 and 3 were done individually.

| Lab | #repos | #events |
| --- | --- | --- |
| 1   | 12     | 223277  |
| 2   | 21     | 90600   |
| 3   | 21     | 262290  |
| All | 54     | 576167  |

The log files are generated on the machine where CodeProber is run. After a session is completed, the students using the tool upload the log files with git to their repositories. We ask students for informed consent to inspect the log files.

Having the students manually upload their log files has the upside that it makes it very clear to them what data we are collecting. However, there is a risk that some log files get missed due to the extra work of uploading via git.

**Self-reported Time**   The course responsible collected anonymous feedback from the students on how much time they spent on the labs. This was done in order to help adjust the labs for future instances of the course. However, this information also fits well with the log analysis, and is therefore used as a supplementary data source.

#### 6.1.2 Data Analysis

Once the course was done, we cloned all the student repositories to a local machine. Then we created a script that traverses all repositories (that had given consent) and processes the log files. The script extracts values from the logs and writes them to csv files that we could import into a spreadsheet. Most extracted pieces of data are quite simple, such as counting how many instances of a certain event type occur. One of the nontrivial points of data being extracted relates to figuring out how much CodeProber is used during the labs.

**Estimating Use of CodeProber**   Based on our observations, CodeProber is often left idle in the background while the developer works on their language tool. When they have updated their tool, or figure out something new they wish to investigate, they bring CodeProber back into the foreground and interact with it. Since CodeProber is mostly left idle, the time difference between the first and last event of a session cannot be used to measure how much the tool is used. Instead, we can detect periods of use by grouping events together if they happen close to each other. We define a mini session as a sequence of events $e_1, e_2, ..., e_n$, where their time differences $|T_{e_{i+1}} - T_{e_i}| \leq \Delta$ for some grouping length $\Delta$. Summarizing the durations of all mini sessions gives us a better approximation of how much CodeProber was used.

**Study of the Use of Property Probes in an Educational Setting**

◼ **Table 5** Approximations of number of times CODEPROBER was used in the program analysis course labs. Computed by grouping events together based on a certain threshold.

| Lab | Average usage counts based on grouping length ($\Delta$) | | | | | | | |
|-----|--------|-------|--------|-------|-------|--------|--------|--------|
|     | 1 sec  | 5 sec | 10 sec | 1 min | 5 min | 15 min | 30 min | 1 hour |
| 1   | 1405.7 | 648.1 | 448.3  | 180.3 | 48.2  | 19.8   | 13.1   | 10.0   |
| 2   | 477.1  | 234.7 | 166.7  | 75.0  | 27.0  | 12.0   | 8.8    | 7.3    |
| 3   | 1195.3 | 565.6 | 408.8  | 188.5 | 60.4  | 26.8   | 17.7   | 12.4   |
| All | 962.8  | 455.3 | 323.4  | 142.5 | 44.8  | 19.5   | 13.2   | 9.9    |

◼ **Table 6** Approximations of the duration CODEPROBER was used in the program analysis course labs. Computed by grouping events together based on a certain threshold.

| Lab | Average usage duration (hours) based on grouping length ($\Delta$) | | | | | | | |
|-----|-------|-------|--------|-------|-------|--------|--------|--------|
|     | 1 sec | 5 sec | 10 sec | 1 min | 5 min | 15 min | 30 min | 1 hour |
| 1   | 0.19  | 0.69  | 1.08   | 3.00  | 7.74  | 11.59  | 13.87  | 16.20  |
| 2   | 0.05  | 0.21  | 0.34   | 0.98  | 2.77  | 4.82   | 5.91   | 7.01   |
| 3   | 0.13  | 0.55  | 0.85   | 2.43  | 7.27  | 12.03  | 15.09  | 18.66  |
| All | 0.11  | 0.45  | 0.70   | 2.00  | 5.62  | 9.13   | 11.25  | 13.58  |

◼ **Table 7** Self-reported durations of labs.

| Lab | Self reported time (hours) | | # Responses |
|-----|---------|--------|-------------|
|     | Average | Median |             |
| 1   | 18.8    | 18     | 13          |
| 2   | 11.1    | 10     | 11          |
| 3   | 38      | 30     | 8           |
| All | 21      | 16.5   | 32          |

## 6.2 Results

In total, 576167 log events were uploaded by the students. The number of events and repositories is presented in Table 4.

### 6.2.1 Amount of CODEPROBER Use

Table 5 shows the average number of mini sessions. Table 6 shows the combined duration of these mini sessions. Both tables show computed values for several grouping lengths ($\Delta$). Note that the log data only contains active interactions within CODE-PROBER. This means that the time spent looking at probe outputs and thinking about what to do is not captured. We believe the data in Table 6 is still quite accurate, but the real numbers may be slightly higher, especially in the columns with lower $\Delta$.

The log data also tells us the number of times the students rebuild their compiler per lab: 103 times on average, and a median of 62. Comparing this with the corresponding number in Table 5 for a grouping of 10 seconds (323.4), we can deduce that about one third of the mini sessions are due to the tools being recompiled. In other words, for each time they rebuild their tool, they perform on average two actions inside CodeProber (creating probes, changing text, etc.). This matches Table 1, which lists liveness from rebuilding the compiler as one of the most used features, with only "Creating probes from text" having a higher usage rate.

If CodeProber is used continuously throughout the labs, then we should be able to estimate how much time students spent on the labs based on how much CodeProber is used. Looking at Table 6 with grouping length set to e.g 1 hour, we can approximate that the average lab takes 13.58 hours to complete. The students self-reported the time they spent on the labs, and this is presented in Table 7. The average self-reported lab duration is 21 hours. Since the self-reporting is anonymous, we do not know if the measured times and self-reported times are from the same individuals. That said, we believe that the numbers are similar enough to indicate that CodeProber is used continuously throughout the labs.

### 6.2.2 Spread of CodeProber Use

Table 8 shows details about the distribution, quantity and duration of the log events that were collected. The three inner tables represent lab 1, 2 and 3 respectively. Each row is a single student or student group for the corresponding lab. The ten center columns show the distribution of the captured log events, normalized across the span of those events. For example, the table shows that students S22,S10 produced 23.8 % of their lab 1 events in the first 10 % of the event span.

The table shows a few patterns in the data. First, across all labs the students produce the largest portion of events in the last 10 % of the time. Second, there is some tendency to use CodeProber more in the beginning of the lab, especially for lab 2 and 3.

We believe the distribution of the log events is due to a combination of the following two things:

1. The students work a moderate amount of time once they get access to the labs. After that they work in quite short bursts, until reaching the deadline of the lab, at which point they put in the largest amount of work.
2. The amount that CodeProber is used is different depending on how far the students have progressed in the labs.

The second explanation is supported by the responses to the question about their workflow in the labs (Section 5.2.4, theme: Continuous use of CodeProber).

We believe the events are more spread out for lab 1 because CodeProber is more useful later in the lab. The lab involves collecting and solving type constraints. Until the students have implemented the base classes required for representing and solving constraints, there is quite little to inspect in CodeProber. For labs 2-3, the data shows some tendency to use CodeProber more in the beginning and at the end of the lab. We interpret this as some students use CodeProber first to build an

# Study of the Use of Property Probes in an Educational Setting

■ **Table 8** Distribution of events per 10-percentile of normalized time during the labs (first table is Lab 1 etc). Columns 1-10, 11-20, etc are the 10-percentiles of time. For example, during Lab 1, the students S22,S10 produced 23.8 % of the events in the first 10 % of the time. The tables also include number of events and number of days between first and last event.

| Students | 1-10 | 11-20 | 21-30 | 31-40 | 41-50 | 51-60 | 61-70 | 71-80 | 81-90 | 91-100 | Events | Days |
|---|---|---|---|---|---|---|---|---|---|---|---|---|
| S22,S10 | 23.8 | 6.0 | 0.9 | 0.0 | 0.0 | 16.2 | 49.9 | 0.0 | 0.0 | 3.3 | 4500 | 17.4 |
| S23,S24 | 4.5 | 7.5 | 0.2 | 0.0 | 0.9 | 0.9 | 12.6 | 10.0 | 30.3 | 33.3 | 469 | 0.1 |
| S14,S8 | 11.5 | 5.3 | 2.2 | 42.6 | 0.0 | 7.9 | 0.0 | 4.9 | 0.0 | 25.5 | 19263 | 31.4 |
| S17,S2 | 0.1 | 8.4 | 4.1 | 36.2 | 0.0 | 11.9 | 15.7 | 17.2 | 0.0 | 6.5 | 22789 | 4.9 |
| S25,S15 | 0.7 | 4.6 | 20.5 | 17.4 | 17.2 | 0.0 | 0.0 | 0.0 | 0.0 | 39.6 | 12574 | 24.2 |
| S19,S6 | 7.7 | 0.0 | 0.0 | 0.0 | 8.8 | 15.2 | 0.0 | 0.0 | 0.0 | 68.3 | 58284 | 15.3 |
| S4,S12 | 0.0 | 0.0 | 0.0 | 7.1 | 20.5 | 0.0 | 0.0 | 18.8 | 0.0 | 53.5 | 14229 | 7.1 |
| S5,S20 | 40.4 | 0.0 | 1.7 | 0.0 | 28.0 | 0.0 | 1.3 | 27.6 | 0.0 | 1.1 | 41336 | 14.0 |
| S21,S7 | 0.2 | 0.0 | 0.0 | 11.9 | 26.5 | 0.0 | 0.0 | 0.0 | 0.0 | 61.4 | 6037 | 11.7 |
| S11,S9 | 0.1 | 0.0 | 17.3 | 27.7 | 3.4 | 19.0 | 32.0 | 0.0 | 0.0 | 0.5 | 12731 | 21.0 |
| S13,S3 | 28.1 | 0.0 | 0.0 | 0.2 | 2.5 | 5.4 | 1.0 | 0.0 | 20.0 | 42.8 | 24340 | 7.1 |
| S18,S16 | 10.2 | 0.0 | 0.0 | 0.0 | 0.0 | 0.0 | 16.8 | 1.9 | 0.0 | 71.2 | 6725 | 6.5 |
| **Average** | 10.6 | 2.6 | 3.9 | 11.9 | 9.0 | 6.4 | 10.8 | 6.7 | 4.2 | 33.9 | 18606 | 13.4 |

| Students | 1-10 | 11-20 | 21-30 | 31-40 | 41-50 | 51-60 | 61-70 | 71-80 | 81-90 | 91-100 | Events | Days |
|---|---|---|---|---|---|---|---|---|---|---|---|---|
| S1 | 0.7 | 0.0 | 0.0 | 0.0 | 0.0 | 0.0 | 25.3 | 38.6 | 7.4 | 28.1 | 4027 | 1.3 |
| S2 | 3.2 | 0.0 | 0.0 | 0.0 | 44.5 | 0.0 | 0.0 | 0.4 | 0.2 | 51.7 | 6920 | 8.1 |
| S3 | 41.9 | 56.7 | 0.0 | 0.0 | 0.0 | 0.0 | 0.0 | 0.0 | 0.0 | 1.5 | 8946 | 7.9 |
| S4 | 82.1 | 0.0 | 10.9 | 0.0 | 0.0 | 0.0 | 0.0 | 0.0 | 0.0 | 6.9 | 3162 | 4.0 |
| S5 | 1.0 | 0.0 | 0.0 | 0.0 | 0.0 | 0.0 | 0.0 | 6.3 | 9.0 | 83.6 | 5144 | 11.7 |
| S6 | 4.1 | 0.0 | 0.0 | 11.8 | 0.0 | 2.9 | 73.5 | 0.0 | 0.0 | 7.8 | 6998 | 5.0 |
| S7 | 24.3 | 32.7 | 8.2 | 0.0 | 0.0 | 0.0 | 0.0 | 0.0 | 0.0 | 34.8 | 4907 | 16.0 |
| S8 | 7.0 | 10.3 | 6.2 | 5.1 | 24.6 | 0.0 | 3.6 | 3.7 | 0.1 | 39.4 | 8736 | 3.3 |
| S9 | 6.4 | 0.0 | 0.0 | 0.0 | 0.0 | 15.2 | 0.0 | 0.0 | 67.7 | 10.7 | 3522 | 8.2 |
| S10 | 61.9 | 34.4 | 0.0 | 0.0 | 0.0 | 0.0 | 0.0 | 0.0 | 0.0 | 3.7 | 4168 | 12.0 |
| S11 | 20.2 | 0.0 | 0.0 | 0.0 | 0.0 | 1.1 | 0.0 | 17.7 | 0.0 | 61.0 | 1953 | 6.7 |
| S12 | 1.9 | 30.3 | 0.0 | 0.0 | 0.0 | 0.0 | 0.0 | 0.0 | 0.0 | 67.8 | 4904 | 1.4 |
| S13 | 1.7 | 0.0 | 19.8 | 8.8 | 17.5 | 5.8 | 0.0 | 10.5 | 13.3 | 22.5 | 2445 | 3.7 |
| S14 | 82.6 | 0.0 | 0.0 | 0.0 | 0.0 | 0.0 | 0.0 | 0.0 | 0.0 | 17.4 | 6540 | 18.9 |
| S15 | 3.6 | 32.1 | 55.4 | 0.0 | 4.4 | 1.7 | 0.0 | 0.0 | 0.0 | 2.8 | 4717 | 23.0 |
| S16 | 24.0 | 0.0 | 0.0 | 0.0 | 0.0 | 0.0 | 0.0 | 0.0 | 36.4 | 39.6 | 984 | 1.2 |
| S17 | 22.2 | 0.0 | 0.0 | 0.0 | 0.0 | 0.0 | 0.0 | 0.0 | 11.1 | 66.7 | 9 | 0.0 |
| S18 | 5.0 | 0.9 | 0.0 | 2.1 | 39.2 | 1.9 | 1.5 | 0.0 | 24.7 | 24.7 | 778 | 0.1 |
| S19 | 17.8 | 32.9 | 0.0 | 0.0 | 13.4 | 12.1 | 0.0 | 0.0 | 0.0 | 23.9 | 4461 | 2.3 |
| S20 | 19.8 | 0.6 | 0.0 | 3.6 | 11.0 | 8.6 | 0.0 | 0.0 | 0.0 | 56.3 | 6217 | 1.3 |
| S21 | 13.3 | 5.0 | 0.0 | 15.9 | 13.3 | 25.0 | 5.9 | 0.0 | 4.3 | 17.2 | 1062 | 0.2 |
| **Average** | 21.2 | 11.2 | 4.8 | 2.3 | 8.0 | 3.5 | 5.2 | 3.7 | 8.3 | 31.8 | 4314 | 6.5 |

| Students | 1-10 | 11-20 | 21-30 | 31-40 | 41-50 | 51-60 | 61-70 | 71-80 | 81-90 | 91-100 | Events | Days |
|---|---|---|---|---|---|---|---|---|---|---|---|---|
| S1 | 2.8 | 7.1 | 21.0 | 0.0 | 3.3 | 11.6 | 12.1 | 5.1 | 21.6 | 15.5 | 12025 | 9.8 |
| S2 | 30.8 | 0.0 | 0.0 | 17.5 | 0.0 | 0.0 | 0.0 | 5.3 | 0.0 | 46.4 | 2825 | 3.1 |
| S3 | 40.1 | 5.1 | 0.0 | 0.0 | 0.0 | 15.1 | 5.9 | 5.3 | 26.8 | 1.6 | 12353 | 7.8 |
| S4 | 0.4 | 0.0 | 0.4 | 9.7 | 64.1 | 24.7 | 0.0 | 0.0 | 0.6 | 0.3 | 5772 | 27.3 |
| S5 | 8.9 | 0.9 | 0.0 | 0.0 | 8.8 | 19.8 | 1.5 | 0.8 | 27.5 | 31.6 | 1413 | 7.0 |
| S6 | 0.3 | 0.5 | 29.8 | 31.2 | 0.9 | 11.1 | 0.0 | 0.6 | 1.0 | 24.7 | 25166 | 10.6 |
| S7 | 0.3 | 2.8 | 0.0 | 39.9 | 0.0 | 0.0 | 0.0 | 15.3 | 23.3 | 18.3 | 8874 | 7.9 |
| S8 | 5.8 | 3.1 | 15.9 | 2.8 | 0.0 | 0.0 | 13.1 | 17.8 | 20.5 | 20.9 | 10349 | 9.9 |
| S9 | 51.4 | 3.8 | 7.8 | 0.0 | 0.0 | 8.6 | 9.4 | 0.0 | 0.0 | 18.9 | 3367 | 9.2 |
| S10 | 1.8 | 5.3 | 0.2 | 2.9 | 56.4 | 0.0 | 19.9 | 0.0 | 0.0 | 13.4 | 12490 | 18.9 |
| S11 | 17.9 | 0.0 | 0.0 | 0.0 | 0.0 | 0.0 | 73.3 | 0.0 | 0.0 | 8.8 | 7422 | 11.8 |
| S12 | 49.7 | 14.5 | 0.0 | 0.0 | 0.0 | 0.0 | 0.0 | 23.6 | 0.0 | 12.1 | 3557 | 3.3 |
| S13 | 1.7 | 0.0 | 0.0 | 0.0 | 5.2 | 5.2 | 8.8 | 13.1 | 31.3 | 34.7 | 7714 | 10.2 |
| S14 | 0.9 | 25.0 | 0.0 | 0.0 | 0.0 | 0.0 | 30.1 | 0.0 | 27.1 | 16.9 | 15544 | 8.2 |
| S15 | 0.2 | 0.0 | 0.0 | 0.6 | 0.0 | 0.8 | 35.9 | 1.1 | 13.3 | 48.0 | 20814 | 12.6 |
| S25 | 8.6 | 0.0 | 0.0 | 0.7 | 22.8 | 0.0 | 15.1 | 21.0 | 0.0 | 31.9 | 3002 | 3.2 |
| S16 | 6.8 | 6.8 | 0.0 | 0.0 | 0.0 | 0.0 | 36.8 | 3.2 | 5.7 | 40.6 | 2032 | 1.2 |
| S26 | 21.2 | 10.1 | 0.0 | 0.0 | 0.0 | 1.0 | 41.7 | 2.4 | 17.8 | 5.8 | 70567 | 15.1 |
| S19 | 6.6 | 3.6 | 39.0 | 0.0 | 0.0 | 11.8 | 4.1 | 0.2 | 19.0 | 15.5 | 3481 | 9.2 |
| S20 | 73.7 | 1.2 | 0.0 | 0.0 | 0.0 | 1.1 | 0.8 | 0.1 | 1.8 | 21.2 | 31940 | 6.1 |
| S21 | 36.3 | 0.0 | 0.0 | 0.0 | 0.0 | 0.0 | 0.0 | 0.0 | 0.0 | 63.7 | 1583 | 7.2 |
| **Average** | 17.5 | 4.3 | 5.4 | 5.0 | 7.7 | 5.3 | 14.7 | 5.5 | 11.3 | 23.4 | 12490 | 9.5 |

understanding of the code they were handed. Once that understanding is in place, the students enter into a phase of mostly developing new features and briefly coming back to CodeProber to check that the most recently developed feature works as expected. Towards the end of the lab, they need to make sure everything is working correctly in order to pass the labs.

## 7 Survey

In this section, we described the method and results of the survey part of the study.

### 7.1 Method

At the end of each course in Lund University, the students are invited to participate in an anonymous survey about the course. Large parts of the survey is standardized by our faculty, but the course responsible is able to insert up to 4 custom questions for their course. These questions come in the form of statements that the student should respond to on a scale of $[-100, +100]$, where $-100$ means "fully disagree" and $100$ means "fully agree".

At the end of the two courses where CodeProber is used, one or more custom questions relating to CodeProber was added. There is a limit of four questions per course instance, and there are other non-CodeProber concerns that need to be surveyed as well. Therefore, we only managed to get one near identical across the two courses:

> **Program Analysis Course Survey**: *CodeProber was effective at helping me discover and understand bugs and omissions in my analysis implementation.*
> **Compilers Course Survey**: *CodeProber was effective at helping me discover and understand bugs in my compiler.*

### 7.2 Results

In total, we have survey responses from three course instances; the program analysis course instance in 2022 and 2024, the compilers course in 2023.

Figure 9 shows the responses from the three course evaluations that included a question about CodeProber. In the program analysis course, 23 of the 29 responses "agree" or "fully agree" with the statement that CodeProber is "effective". In the compilers course, the corresponding numbers are 15 out of 22.

The more positive response in the program analysis course can be explained by a few factors. For one, CodeProber is introduced from the very first lab, so it becomes a more integral part of the students' workflow early on. In the compilers course, the students complete $\sim 2.5$ labs before seeing CodeProber, so by that point they have gotten used to working with test cases.

Another factor to explain the difference may be the nature of the lab assignments. In the compilers course the labs vary significantly. The tasks include parsing, analyzing, interpreting and generating code from source code. We have anecdotal evidence from

**Study of the Use of Property Probes in an Educational Setting**

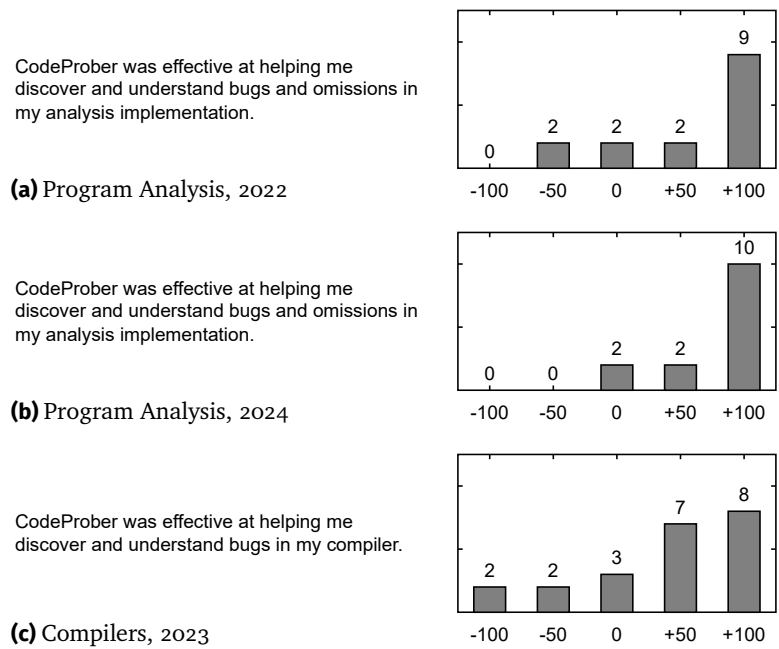

■ **Figure 9** Results from course evaluations in three course instances. X-axis goes from −100, meaning "fully disagree", up to +100, meaning "fully agree". Y-axis is the number of individual responses.

talking with students that CODEPROBER is most used and most appreciated by the students in the analysis lab. The program analysis course labs exclusively consist of analysis-related tasks, and CODEPROBER may therefore be an overall better fit for that course. Another difference is that the students are given an existing compiler in the program analysis course that they extend, rather than building it from scratch (like in the compiler course). Thus, the need to build an understanding of a given codebase is bigger in the program analysis course.

## 8  Summary of Results

In this study, we have three sources of data, whose background and individual results are presented in Sections 5, 6 and 7 respectively. In this section, we combine all the individual results to answer our research questions.

### 8.1  RQ1 What is the user experience of using CODEPROBER in an educational setting?

**RQ1** can be answered by the interviews and course evaluations. Both paint a positive picture of CODEPROBER. The students found the liveness and exploratory nature of CODEPROBER useful. Being able to explore their language tool without having to add print statements or set a breakpoint enables quick and convenient usage. The course evaluations also confirm the usefulness of CODEPROBER. While the course evaluations are not as extensive as the interviews, they are anonymous, allowing students to voice their opinions without social pressure. Since the survey answers (despite their limited scope) are positive, we are more confident in the interview results.

The most common point of negative feedback was related to freezes and crashes. The way CodeProber was used in the 2024 instance of the program analysis course highlighted a few bugs we had not seen before. We plan on improving this for future course instances.

In the interviews, we asked the students questions regarding the usefulness of Code-Prober and how much they "like" using it. We believe both aspects are important to the overall user experience. For example, while most (N=10) interviewees agree that test cases are useful, a majority (N=6) of interviewees also mention that they dislike writing test cases, which can lead to less automated testing overall. The score in Table 3 shows that interviewees, on average, "strongly like" using CodeProber.

In summary, we consider the answer to **RQ1** to be that the students find Code-Prober to be a useful tool that is enjoyable to use, despite some technical issues.

### 8.2 RQ2 How is CodeProber used during the development of compilers and static analysis tools in an educational setting?

**RQ2** can be answered by the results from the interviews and analysis of log files. From the interviews, we find that CodeProber is used to varying degrees throughout the entire labs. This is confirmed by the log files, as there are a number of log events throughout the entire lab series. Almost half of the events occur in the first and last 10 % of the labs. We interpret this as some students use CodeProber first to build an understanding of the handout code and later on to verify that everything works as expected.

In terms of features, students mainly focused on standard probes, squiggly lines and liveness, as shown in Table 1. The other features are less used, either due to lack of need or because they did not know those features existed. Whether this means that the other features are worthwhile is hard to say, as more experienced developers may have different usage patterns. One thing we can say however is that feature discoverability within CodeProber may need some improvement. For example, a majority (N=7) of interviewees remarked that *search probes* seem useful after we demonstrated it to them, but they had no idea that the feature existed.

In summary, we consider the answer to **RQ2** to be that the students made continuous use of CodeProber, and they mainly rely on standard probes, squiggly lines and liveness.

### 8.3 RQ3 How does the use of CodeProber compare to other tools used by students during the development process (debuggers, test cases, print-statements, AI, etc.)?

**RQ3** can be answered by results from the interviews. Table 2 shows that students use CodeProber more than any other tool in the labs. Test cases and print debugging are quite close to each other in a shared second place, and breakpoint/step debugging and AI assistants are last.

**Study of the Use of Property Probes in an Educational Setting**

The reason for not using those other tools can in part be explained by the developer experience of working with JastAdd/RAGs. Some attributes in RAGs can have their results cached, and others may evaluate themselves multiple times until a fixpoint is achieved. This means that when an attribute with a print statement is invoked, the print may not execute at all, or it might execute many times, depending on what has been cached in the AST so far. This hurts the usefulness of print statements. Similarly, a breakpoint/step debugger would have to step through the code that handles caching and fixpoint iteration, which is likely not what the developer is interested in. So while we found in the interviews that CodeProber is used a lot more than those other techniques, it is not necessarily because the experience is so much better. It can be because the experience of print debugging and breakpoint/step-debugging is in general worse when working with RAGs, and CodeProber is able to fill that role instead.

Note that test cases are different: we believe that they are generally as useful for RAGs as in general software development. Here, we believe the problem is that it is too quick and convenient to verify that something works in CodeProber, so students do not want to spend the extra time in creating a proper test case. CodeProber does not currently support any form of regression testing, so the reduced use of test cases is not a desired outcome. To help combat this, we plan on adding test support to CodeProber, i.e. the ability to save a probe as a test case that can later be run in for example in JUnit. That would reduce the barrier of creating a test case to a few clicks. In addition, we believe this problem will likely naturally disappear in larger projects, where manually verifying all functionality simply is not practical. There the developers will have a stronger motivation to write tests.

In summary, we consider the answer to **RQ3** to be that the students in our study used CodeProber to partially replace print debugging, breakpoint/step-debuggers and test cases. We hypothesize that this is in part because the challenges of debugging RAGs are not handled as naturally by those traditional tools. We cannot say much about AI because none of our survey participants used it very much, neither in the course nor in general software development.

**8.4 Threats to Validity**

**Internal Validity** There is risk of bias in the findings from the interviews due to participant responder bias [4]. We have interacted with a majority of these students before in some capacity, either as teaching assistants or supervisors in various courses. The main author of this paper was a teaching assistant in the program analysis course. The students know that we are developing CodeProber, and therefore they may want to be "kind" and mostly say positive in the interviews. We tried to mitigate this by explicitly asking for feedback on things that do not work well, and by having the person they knew the least well lead the interviews. This risk of bias is also why we spent time on analyzing log files and course evaluations, as these are anonymized and can help verify whether the interviews were overly positive or not.

When designing surveys, acquiescence bias needs to be considered, that is, the tendency for respondents to agree with statements as they are presented. The course survey in our study only contained a single statement relating to CodeProber, due to the size restrictions mentioned in Section 7. The statement is phrased positively, which may positively skew the responses we got. A common method to prevent this bias is to add both positive and negative versions of the same statement [28, 38]. However, due to the limited total number of statements that could be added to the survey (4), and the fact that there were several aspects of the courses that should be surveyed (not just about CodeProber), we could only add one common statement across both courses. Despite the potential for bias, we still believe the survey is valuable due to its anonymity.

The number of students in the interviews is quite low (9), and is a potential threat to the interview findings. However, we did notice data saturation in the interview responses, and we believe that additional interviews would not significantly affect the results.

**External Validity**   This study was conducted on a limited set of students in one specific educational setting. This limits the generalizability of the results. Students from different universities, educational settings or cultural contexts may respond differently to a tool like CodeProber. CodeProber is mainly created to help develop JastAdd-based tools, as was the case in this study. This means that the results may not be generalizable to other students unless they also use JastAdd, or a similar RAG-based language tool stack. Still, we believe that it would be possible to reproduce the results of the study in a different educational setting, provided that the student group has a similar educational background and use a similar tool stack.

Due to the focus on an educational setting, the results may not be generalizable to other contexts, e.g., industry, hobbyists, etc. However, we still believe that CodeProber can be useful for other language tool developers who use a similar tool stack.

## 9  Related Work

In this section we present related work for four different aspects of this study: liveness in development tools, user studies, language tool development, and use of debugging tools.

### 9.1 Liveness

Many live development tools have similar overall designs. They have a text area where the developer can input code. When changes are made, the tool immediately runs the code, extracts runtime information and displays the information back to the developer [5, 12, 22, 23, 26]. A significant difference between CodeProber and these other tools is that the target audience for CodeProber is language tool developers, whereas the other tools target programmers in general. The users of CodeProber are usually interested in the behavior of the language tool for a particular input code rather than the behavior of the input code. Like other live development tools, the values displayed in CODEPROBER are computed by the underlying language tool for a given input code. One difference is that the values in CODEPROBER are usually static information about the input code (e.g., the type of an expression) rather than its dynamic behavior. However, the underlying language tool can provide properties that compute the input code's dynamic behavior and display it in CODEPROBER. It would be interesting future work to improve the visualization of such dynamic information, for example, by integrating *projection boxes* (see below) in CODEPROBER. CODEPROBER supports changing the underlying language tool, and this also changes which language is supported in the text area and what properties that are available. In contrast, other tools generally only support one language. We will discuss some of the other tools here in more detail, and how they compare to CODEPROBER.

Lerner [23] presented VERSABOX, a Python editor that displays runtime values of variables in floating windows (*projection boxes*) next to the code. The boxes move around so that the information for the currently focused line is displayed most prominently. In CODEPROBER, the probes are used to select entry points into the underlying language tool. This lets the developer pick which subset of possible information they want to see at any given time. In VERSABOX, the full program is always executed, and in the paper they mention that information overload is a potential problem. We believe CODEPROBER's design scales better when developing larger language tools, both in terms of usability and performance. It would be interesting future work to explore integration of projection boxes into CODEPROBER. The probes could be used as entry points for collecting runtime information of the underlying language tool.

McDirmid [26] presented YINYANG, a code editor and language that supports live "probing" of expressions. The language supports prefixing any expression with an @-sign. This has no impact on the semantics of the expression, but it causes the runtime value of the expression to be rendered in a small box ("probe") inside the code editor. The ability to select which expression to probe via the @-sign is similar to how CODEPROBER only displays information once a probe has been created. This helps reduce information overload. However, we believe YINYANG has similar scaling issues as VERSABOX when developing non-trivial language tools. For example, the developer may want to inspect the type of a specific expression inside a large input text. With YINYANG, they would have to 1) parse the input text, 2) programmatically locate the node representing the expression inside the parsed tree, and 3) invoke a "type" function (and possibly its dependencies) with an @ decorator. In CODEPROBER, the developer would right-click the expression and create a probe for "type".

**9.2 User Studies**

There are several user studies that investigate the effect of liveness in development tools. A common theme [9, 16, 22, 23, 25, 39] is to perform controlled experiments with beginners (e.g., students), and tasks that should be solved during the experiment, with or without live feedback. In contrast, the students in our study use CODEPROBER for several weeks, and work in a codebase of several thousand lines of code. We believe this makes it possible to make interesting observations of how CODEPROBER is used to develop larger, real-world-like codebases. Nonetheless, it could be interesting as future work to perform a controlled experiment with CodeProber too. In this section we discuss a few of the related studies in more detail.

Hundhausen et al. [16] studied the effect using development environments with three levels of feedback: 1) No feedback is given, 2) Feedback is given after a button is pressed, and 3) Live feedback is always given. They found that any form of feedback is better than none. However, they found no significant difference between continuous live feedback and feedback that was given after a button press. They suggest that continuous live feedback may sometimes be a distraction, and the button allows the developer to wait until they are ready to take advantage of the feedback. CodeProber combines liveness and letting the developer choose when to receive feedback. By default, CodeProber gives no feedback to the user. Only when the user has created a probe will some form of feedback be given. In addition, changing the specification for the underlying language tool has no immediate impact, the developer must compile the underlying language tool for changes to appear inside CodeProber. Creating probes and compiling the underlying language tool can be compared to the button in Hundhausen et al.'s study. They enable the developer to get live feedback, but only when they are ready for it. Once a probe is created, then the feedback will be given continuously when the input text is changed, like scenario 3 in Hundhausen et al.'s study.

Kramer et al. [22] extended the JavaScript IDE Brackets[4] with a view that displays runtime values of the code to the side, similar to the work on projection boxes discussed above. They also performed a user study where they compared the performance of developers based on if they received live feedback inside the editor or not. Interestingly, they did notice some significant workflow differences. Developers without their extension (without live feedback) tended to solve tasks in two phases: first they implemented most of the functionality, and then they spent time on making sure everything works. The developers with live feedback instead tended to continuously fix issues as they wrote the code. In our study, we did not have a control group that worked without CodeProber, so we can not observe this difference directly. However, some interviewees did hint towards this iterative develop-fix workflow (Section 5.2.4, theme: Continuous use of CodeProber).

Rein et al. [32] extended a Squeak/Smalltalk environment with a *"cross-cutting perspective"* to help the developer filter and visualize execution traces. This perspective was evaluated in an exploratory user study with 7 students. The interviews were each 2.5 hours long. First, the interview subjects worked to solve some tasks using the new perspective for 1.5 hours. Then for the remaining time they were asked about their experiences. Live task solving like this could have let us make some more detailed observations about exactly how CodeProber is used.

---

[4] https://brackets.io/. Accessed 2025-02-04.

### 9.3 Language Tooling Development

Language workbenches are tools that "supports the efficient definition, reuse and composition of languages and their IDEs" [8]. Language workbenches are often able to provide liveness features, in part because they control and tightly integrate parsing, semantic specification, testing, and more. For example, Gabriël et al. added incremental compilation of the grammar in Spoofax [21]. This enabled grammar changes to immediately show updated parse trees and/or syntax errors within Spoofax. Dubroy et al. [5] created Ohm, a workbench that lets the developer specify a grammar, examples (≃test cases) and semantic actions. The actions behave similarly to synthesized attributes in Reference Attribute Grammars [14], except they are specified on the concrete syntax tree rather than the abstract one. Whenever grammar, examples or actions are modified, live feedback is given to the user. The tight integration inside a workbench enables some features they provide, but it can also sometimes be a limitation. For example, if one wanted to use Spoofax or Ohm with a custom parser, then this would likely require forking and modifying the respective tool's source code. JastAdd helps the language tool developer with specifying semantics, but does not have any opinions regarding parsing or debugging. The goal of CodeProber is to provide debugging functionality for JastAdd and tools similar to it. By itself, CodeProber is not a language workbench. However, by combining a parser generator, JastAdd and CodeProber, it is possible to get an experience similar to one provided by language workbenches.

There exists several other tools that enable some form of AST exploration. Some examples include Noosa [35], DrAST [24], Aki [17], and EvDebugger [33]. None of these tools have the concept of probes that are updated after changes to the source text. They also focus much of their interaction on the AST. CodeProber on the other hand handles most user interactions in terms of source code, which we believe scales better for more complex language tools and when exploring larger input texts. For example, we think that displaying and navigating an AST view for 500 lines of Java code is non-trivial, while rendering and navigating 500 lines of text is simpler. These tools are discussed in more detail in [2].

### 9.4 Debugging

CodeProber can aid in the process of debugging, but it is not a traditional debugger by itself. Part of the goal of this study is to investigate if, and to what degree, the need for debugging can be met by CodeProber. This section presents studies on the usage (or lack thereof) of traditional debuggers.

Ko et al. [20] performed a survey to investigate which barriers and factors prevent students from making more use of debuggers. They did this with an online survey where 73 students participated. They found that the lack of focus on debugger usage in academic courses is one of the main reasons. Also, the complexity of debuggers high initial learning curve was one reason. In our interviews a majority (N=8) of interviewees also mentioned that a downside of debuggers is that they can be quite difficult to set up and run.

Beller et al. [3] performed a mixed-method study to determine "how developers debug software problems in the real world". They performed an online survey where 458 developers participated, and performed automated data collection from IDEs of 108 people. They found that "developers spend surprisingly little time in the debugger; only 13 % of their total development time on average". Many developers preferred using print debugging instead. The reasons they found for this behavior include the complexity of modern debuggers. Another reason is experience – they found that more experienced developers tended to use debuggers slightly more. In the context of language tools, we believe that the design of CodeProber solves some problems of complexity found with traditional debuggers, as discussed in Section 2. That is, finding the AST node of interest during debugging is non-trivial with traditional debuggers, but is one of the primary features of CodeProber. The fact that our students seem to use CodeProber more than debuggers, test cases and print debugging strengthens this claim. But even with CodeProber, we believe that debuggers are still very useful, especially for more complex bugs.

## 10 Conclusions

We have presented a mixed method study of the experience of using CodeProber in an educational setting. Our findings show that the students find CodeProber to be useful, and they make continuous use of it during the course labs. One of the most used features is its liveness, e.g. the ability to nearly instantly respond to changes. We found that CodeProber to some extent replaces existing tools and techniques like test cases, breakpoint/step-debugger, etc. We hypothesize that this is in part because the nature of RAGs does not work so well with these more general tools, so CodeProber is able to better meet this demand. The reduced use of test cases can be seen as a negative outcome, as CodeProber does not support automatic regression testing. For this reason, we are interested in adding support for testing inside CodeProber in the future.

Further research is needed to explore how CodeProber or a tool like it would be perceived by people outside an educational setting, for example industry professionals. There is also interesting future research in trying to apply it to completely different domains. We have used CodeProber exclusively for language tooling, but in theory any program that performs computations on a tree structure is a possible target. For example, many models for building UI are tree structures. Scene graphs (often used in games) and the Document Object Model ("DOM", used in browsers) fit into this category, and are therefore possible future targets for CodeProber.

**Acknowledgements**   We want to thank all the participants in this study, Christoph Reichenbach for integrating CodeProber into the program analysis course and Görel Hedin for integrating CodeProber into the compilers course. We would also like to thank the reviewers for helping us improve the paper.

The authors would further like to thank the following funders who partly funded this work: the Swedish strategic research environment ELLIIT, the Swedish Foundation for Strategic Research (grant nbr. FFL18-0231), the Swedish Research Council (grant nbr. 2019-05658), and the Wallenberg AI, Autonomous Systems and Software Program (WASP) funded by the Knut and Alice Wallenberg Foundation.

**Data-Availability Statement**    The interview design, coding scheme, results from log file analysis, etc., are available on Zenodo [1]. The handout code for the program analysis labs is also available on Zenodo [31]. The source code of CodeProber is available on https://github.com/lu-cs-sde/codeprober.

## A   Interview Design

**Figure 10** The script used during interviews, page 1.

---

# Interview structure

## 0. Physical setup

- The interview happens in Niklas' room
- One "demo computer" with VS Code and CodeProber running, in case we/the interviewee wants to show something
  - This computer will also run audio recording the whole time
  - Process for getting the computer ready:
    - Open quicktime, prepare for starting audio recording (File->New Audio Recording, don't press start yet)
    - Open the directory `lab-2` in Visual Studio Code
      - Same as `lab-2` in artifact https://doi.org/10.5281/zenodo.13380279, except for a small modification to make at least 1 "report" show up in CodeProber.
    - Run `./gradlew clean jar`
    - Run `./code-prober.sh examples/hw2-task-2-0.teal`
    - Click the link `http://localhost:8000/` to open in browser of choice, preferably Chrome or Firefox (there may be issues in Safari)
    - Close all other windows to avoid distraction (only keep Quicktime, VS Code and browser).
- One phone, running audio recording as well, laying on the table
- One note-taking computer, to be used by the interviewer that isn't talking.
- Some cans of drinks (soda, water, ..) and cookies
- A set of printed papers for each test subject:
  - Informed consent form (2 copies, one to sign and one to take home)
  - The tables to be filled in (Table 1, 2 and 3 below)
- Pens

## 1. Introduction

### 1.1. Background

- Offer a drink and cookies
- Hello, welcome. Introduce Niklas and Anton. Niklas will ask questions and Anton will take notes.
- Go over overview and purpose of study:
  - CodeProber is an active research project
  - We are curious of how you use CodeProber
    - We try different solutions
    - We would like to know what work and what doesn't work
    - We are happy to get honest feedback, including flaws
    - This helps us better understand the tool and how it can be improved

**Study of the Use of Property Probes in an Educational Setting**

■ **Figure 11** The script used during interviews, page 2.

---

- Research questions:
    - How is CodeProber used during the development of compilers and static analysis tools?
    - What is the user perception of CodeProber? Things that you like/don't like.
    - How does CodeProber compare to other tools during the development process (e.g debuggers, test cases, print-statements)?
- Interview Information:
    - We will record for transcription purposes. These transcriptions will be stored locally on our devices (accessed by Niklas/Anton).
    - Anonymized results will be discussed in the research team for this study (Anton, Niklas, Emma Söderberg).
    - Anonymized results from interviews may be included in a publication
    - You can withdraw from the study within 1 month of the interview
- Ask participant to sign Informed Consent Form
- Start audio recording.

## 1.2. Warmup

**Hand them this form on a printed piece of paper:** Table I: Background Information. Ask them to fill the form. Note: 1 question require a longer answer, so it should be answered verbally. Also ask:

- Experience of compiler development/program analysis beyond the compiler and program analysis course?

Briefly describe the rest of the interview:

- We'll begin by talking about CodeProber specifically
- Then talk about other development tools, and how they all fit together

## 2. Main

## 2.1. Intro to CodeProber

**Showing CodeProber**

- Present one of our laptops. It has VS Code and CodeProber running on it.
- Exercise 2 - dataflow analysis. Null-pointer- and dead assignment analysis
- Please show us and think aloud:
    - Please open a probe showing `Program.reports`
    - Niklas makes changes to remove the null pointer bug (illustrating liveness)
    - Please open an AST view / AST probe for the function called `f`

**Features**

- ***Hand them this table on a printed piece of paper (2 sided):*** Table II: CodeProber Features
- CodeProber supports a number of different features.
- For each one of the following, please fill in if/how much you use it.

**Figure 12** The script used during interviews, page 3.

- Please think aloud
- (If they don't recognise a feature, show it on the laptop)

### 2.2. Workflow

Lets pretend that you are working on one of the labs in the program analysis course. On a high level, what are you spending your time on? Where do you look for information, what editor do you write code in, how do you debug problems, etc. In other words, please describe your workflow. You can describe it just using words, or use the laptop as a reference.

- (Wait a while while they answer, then add the questions below)
- Can you estimate how large portion of your time you spend interacting with the things you mentioned? For example, X% writing code, Y% in CodeProber, Z% reading lecture slides, etc.
- Do you ever look at the code generated by JastAdd?
- If/when you work with test cases, do you prefer writing them early (e.g "Test-Driven Development", TDD), late, or a mix of both?
- Is there any difference in the workflow based on how far along you are in the lab? For example:
  - In the beginning, in the middle, and in the end?
- If you took the compiler course, was there any difference in your workflow for those labs?

### 2.3. Comparison to other development tools

When working in a codebase, you almost inevitably run into problems. When you do, there are several different tools or techniques you can use to overcome the problem.
We have chosen to focus on 5.

*Hand them this table on a printed piece of paper:*

- Table III: Development Tools
- First table is about how much you use a tool/technique
- Second table is about how you like using the tool/technique.
- Please fill in the tables and think aloud

For each technique/tool also discuss:

- How much they are used when you develop software in general
- Its main strengths and weaknesses

Do you feel there is a scenario or tool we have missed above? If so, please add it! (they might mention code review or pair programming here perhaps)

### 2.4. Perception of CodeProber

- Back to CodeProber
- What do you like most with CodeProber? Why?
- What do you like least with CodeProber? Why?
- What feature(s) would you want to add to CodeProber?

**Figure 13** The script used during interviews, page 4.

> If you took the compiler course, did you notice any differences between CodeProber in the compiler vs program analysis course?
>
> - If yes, any thoughts on the difference?
>   (Main difference from their perspective is probably the predefined minimised probes)
>
> ### 2.5. Thoughts on labs
>
> In the program analysis course labs you have had access to CodeProber every lab so far.
>
> - What do you think about the labs?
> - Was your workflow different in the different labs? For example, did you use test cases more or less in certain labs, etc.
>   - In case they need reminder of the different labs, they are:
>     1. JastAdd intro
>     2. Type Inference
>     3. Dead assignment analysis
>     4. Interval & array bounds analysis
>
> If you took the compiler course, then the similar question as before.
>
> - What do you think about the labs?
> - Did you use CodeProber more or less in any of the labs?
>   - In case they need reminder of the different labs, they are:
>     1. Scanning & Parsing (LL1)
>     2. Scanning & Parsing (LR)
>     3. Visitor pattern, basic properties
>     4. Semantic analysis (names & types)
>     5. Interpretation
>     6. Code generation
>
> ### 3. Cooldown
>
> - Anything you would like to add to this interview?
>   - Topics we have missed/not spent enough time on
> - Do you have any questions to us?
>
> ### 4. Closing
>
> - Thank you for participating, here is your Coffee mug

## B  Background Information Form

**Figure 14** The first form filled in by the interviewees.

### Table I: Background Information

| | |
|---|---|
| Which program are you in? (D, C, ..) | |
| Which year are you in at LTH? | |
| How many years have you been programming? Free time counts! | |
| How would you compare your programming skills to your classmates? | Much Worse / Worse / Identical / Better / Much Better |
| Did you take the compiler course? | |
| What made you apply to the program analysis course (and compiler course, if applicable) | *(answer verbally)* |

## C CODEPROBER Feature Form

**Figure 15** The second form filled in by the interviewees, page 1.

| Table II: CodeProber Features ||
|---|---|
| **Feature** | **How much you use it**<br>x = Haven't heard of it before<br>0 = Never<br>1 = Very Rarely<br>2 = Rarely<br>3 = Sometimes<br>4 = Often<br>5 = Very Often |
| Creating probes by right clicking in the text 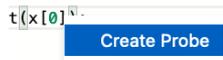 | |
| Creating probes by clicking on node references in the output of a probe 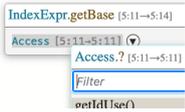 | |
| Inspecting values of attributes/properties in the tiny floating windows, also known as "probes" 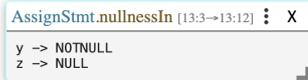 | |
| Hovering AST node references in a probe to see where they are in the text. | |
| Inspecting the AST (the graphical view) 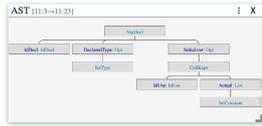 | |
| Nested probes, i.e clicking the downwards-facing triangle ('▼') next to an AST node reference to create a probe inside a probe. 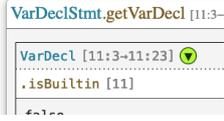 | |
| Minimizing probes to- and opening probes from- the minimized area at the top of the screen 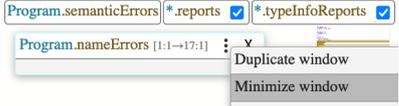 | |

▉ **Figure 16** The second form filled in by the interviewees, page 2.

| Feature | How much you use it<br>x = Haven't heard of it before<br>0 = Never<br>1 = Very Rarely<br>2 = Rarely<br>3 = Sometimes<br>4 = Often<br>5 = Very Often |
|---|---|
| Looking at/drawing arrows, showing e.g the control-flow graph 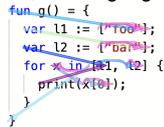 | |
| Looking at/conveying information in squiggly lines 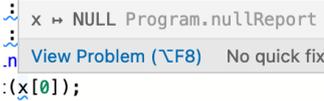 | |
| Changing the text inside CodeProber and seeing the probes update live | |
| Re-building your compiler and seeing the probes update live | |
| Stopping a long-running probe via the stop button in the UI 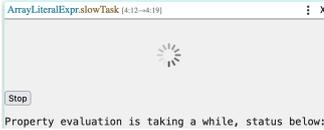 | |
| Search probes, like "*.nullnessValue" 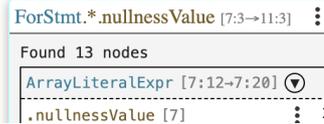 | |
| Tracing 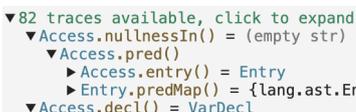 | |
| Other feature (if there is something you feel is missing above) | |

**Study of the Use of Property Probes in an Educational Setting**

# D  Tool/Scenario Form

■ **Figure 17** The third form filled in by the interviewees.

**Table III: Development Tools/Techniques**
During the development of compilers or static analyzers

How much do you **use** each tool/technique

| Scenario | Printf debugging | Breakpoint/step debugging | Test cases | CodeProber | AI (Copilot / ChatGPT / similar) |
|---|---|---|---|---|---|
| **Developing a new feature** | | | | | |
| **Building understanding of a codebase** *Could be code you wrote yourself some time ago, or somebody else's code* | | | | | |
| **Fixing a bug** | | | | | |

x = Haven't heard of it before
0 = Never
1 = Very Rarely
2 = Rarely
3 = Sometimes
4 = Often
5 = Very Often

How much do you **like** using each tool/technique

| | Printf debugging | Breakpoint/step debugging | Test cases | CodeProber | AI (Copilot / ChatGPT / similar) |
|---|---|---|---|---|---|
| | | | | | |
| | | | | | |

x = Haven't heard of it before
1 = Strongly dislike
2 = Dislike
3 = Neutral
4 = Like
5 = Strongly like

Study of the Use of Property Probes in an Educational Setting

## About the authors

**Anton Risberg Alaküla** is a PhD student in Computer Science at Lund University, Sweden. His research interests include programming languages and cloud computing. He received his MSc. in Computer Science and Engineering in 2014 from Lund University, and spent 7 years in industry before starting his PhD. Contact Anton at anton.risberg_alakula@cs.lth.se
https://orcid.org/0000-0003-0814-3367

**Niklas Fors** is an Associate Senior Lecturer at Lund University, Sweden. His primary research interests are tooling for software languages, program analysis and reference attribute grammars. In his PhD thesis, he presented a feature-based diagram language Bloqqi for improving code reuse in automation programming. He also teaches introductory programming for engineering students. Contact Niklas at niklas.fors@cs.lth.se
https://orcid.org/0000-0003-2714-9457

**Emma Söderberg** is an Associate Professor at Lund University, Sweden. Her research interests are in the intersection of programming language and tool construction, empirical software engineering, and human-computer interaction, with special interest in the programmer experience. She has a background as a software engineer at Google and her PhD thesis work was on generation of semantic editors using reference attribute grammars. Contact Emma at emma.soderberg@cs.lth.se
https://orcid.org/0000-0001-7966-4560